\documentclass[letterpaper]{article} % DO NOT CHANGE THIS
\usepackage{aaai2026} 

\usepackage{times}  % DO NOT CHANGE THIS
\usepackage{helvet}  % DO NOT CHANGE THIS
\usepackage{courier}  % DO NOT CHANGE THIS
\usepackage[hyphens]{url}  % DO NOT CHANGE THIS
\usepackage{graphicx} % DO NOT CHANGE THIS
\usepackage{natbib}  % DO NOT CHANGE THIS AND DO NOT ADD ANY OPTIONS TO IT
\usepackage{caption} % DO NOT CHANGE THIS AND DO NOT ADD ANY OPTIONS TO IT
\usepackage{algorithm}
\usepackage{algorithmic}
\usepackage{xcolor}

\newcommand{\answerYes}[1]{\textcolor{blue}{#1}} 
 
\newcommand{\answerNA}[1]{\textcolor{gray}{#1}}

\usepackage{listings}
\usepackage[most]{tcolorbox}

\lstdefinestyle{prompt}{
  basicstyle=\ttfamily\tiny,
  frame=single,
  breaklines=true,
  showstringspaces=false,
  captionpos=b,
  numbers=none,
  aboveskip=10pt, 
  belowskip=10pt,
}

\newcommand{\xhdr}[1]{\vspace{1.7mm}\noindent{{\bf #1.}}}

\usepackage{newfloat}
\usepackage{listings}
\DeclareCaptionStyle{ruled}{labelfont=normalfont,labelsep=colon,strut=off} % DO NOT CHANGE THIS
\floatstyle{ruled}
\newfloat{listing}{tb}{lst}{}
\floatname{listing}{Listing}
\usepackage{amsmath}

\title{Persuasion in Online Conversations Is Associated with Alignment in Expressed Human Values}
\author{
    Bhavesh Vuyyuru,
    Farnaz Jahanbakhsh
}
\affiliations{
    University of Michigan\\
    bvuyyuru@umich.edu, farnaz@umich.edu
}

\usepackage{bibentry}
\begin{document}

\maketitle

\begin{abstract}

Online disagreements often fail to produce understanding, instead reinforcing existing positions or escalating conflict. Prior work on predictors of successful persuasion in online discourse has largely focused on surface features such as linguistic style or conversational structure, leaving open the role of underlying principles or concerns that participants bring to an interaction. In this paper, we investigate how the expression and alignment of human values in back-and-forth online discussions relate to persuasion. Using data from Reddit's ChangeMyView subreddit, where successful persuasion is explicitly signaled through the awarding of deltas, we analyze one-on-one exchanges and characterize participants' value expression by drawing from Schwartz's Refined Theory of Basic Human Values. We find that successful persuasion is associated with two complementary processes: pre-existing compatibility between participants' value priorities even before the exchange happens, and the emergence of value alignment over the course of a conversation. At the same time, successful persuasion does not depend on commenters making large departures from their typical value expression patterns. We discuss implications of our findings for the design of online social platforms that aim to support constructive engagement across disagreement.

\end{abstract}


\section{Introduction}

Online discussions are a central venue for public deliberation, exchange of diverse perspectives, and collective sensemaking~\cite{dahlberg2001internet}. The consequences of these interactions are not confined to online spaces; how people argue, persuade, and disengage can shape beliefs, relationships, and behaviors offline as well~\cite{bond201261, bavel2020using}. Yet many such conversations fail to produce understanding or progress, instead devolving into hostility, disengagement, or participants talking past each other~\cite{cheng2017anyone, cheng2015antisocial}. These breakdowns motivate a better understanding of when online disagreements in which participants advance competing claims and attempt to persuade one another, lead participants to genuinely reconsider their views, as opposed to stalling, escalating, or simply reinforcing existing positions.

Prior work has examined factors associated with outcomes in online debate, including linguistic framing, emotional tone, and conversational structure \citep{tan2016winning, ta2022inclusive, althoff2014ask}. Much of this research focuses on properties of discourse and interaction dynamics at the level of surface linguistic features or thread structure. While these features capture important aspects of how arguments are expressed, they operate primarily at the level of surface form and interactional structure, rather than the underlying concerns or principles that shape what participants find compelling in the first place.

Human values offer a principled way to operationalize such underlying priorities. Values are relatively stable, trans-situational principles that people draw on when interpreting information and making trade-offs across contexts \citep{schwartz2012refining}. As an analytic lens, values offer a way to characterize patterns of expression in argumentation that is less tied to specific wording or momentary affect. In many debates, participants justify their positions by appealing---implicitly or explicitly---to underlying values, and arguments may resonate not only because of their logical structure but because they foreground priorities that matter to the listener~\cite{long1983role}.

\begin{figure}[t]
    \centering
    \includegraphics[width=\linewidth]{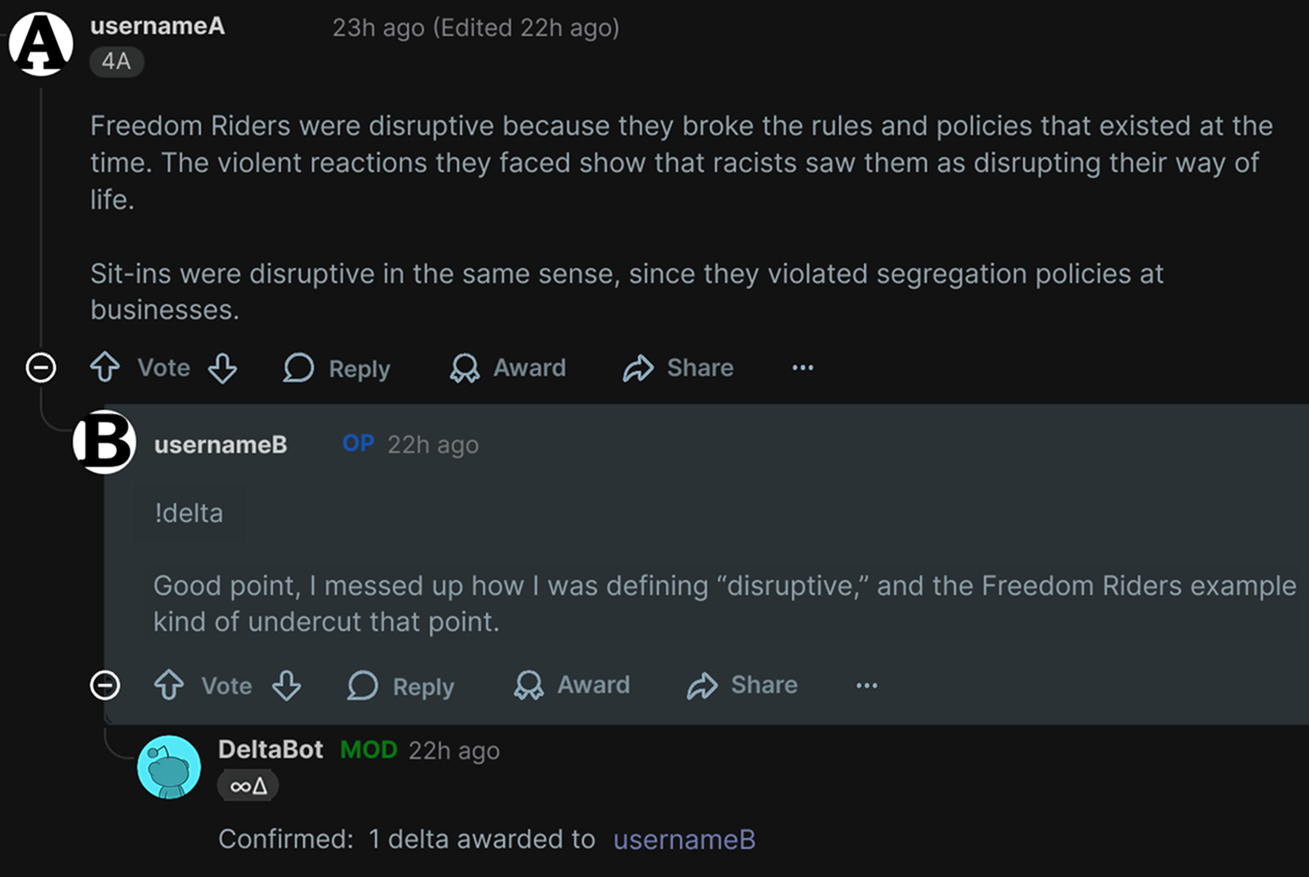}
    \caption{Example exchange from the CMV subreddit in which a commenter successfully persuades the original poster (OP), resulting in a delta award given by the OP to the commenter.}
    \label{fig:convo_example}
\end{figure}

How might we empirically examine the role of values in online argumentation? Addressing this question requires two components: a setting in which persuasive outcomes are observable, and a way to characterize the values participants express through their arguments.
Studying debate and persuasion dynamics in the wild is challenging as moments of successful persuasion are rarely observable. Reddit's ChangeMyView (CMV) subreddit offers a rare opportunity in this regard. On CMV, users post opinions they are open to reconsidering, and commenters attempt to persuade them. When an original poster (OP) feels that a comment has changed their view, they explicitly award a ``delta'', providing a direct signal of successful persuasion (See Figure~\ref{fig:convo_example}). This explicit outcome allows us to study whether successful persuasion relates to the values expressed by participants over the course of a conversation.

\begin{figure}[t]
    \centering
    \includegraphics[width=\linewidth]{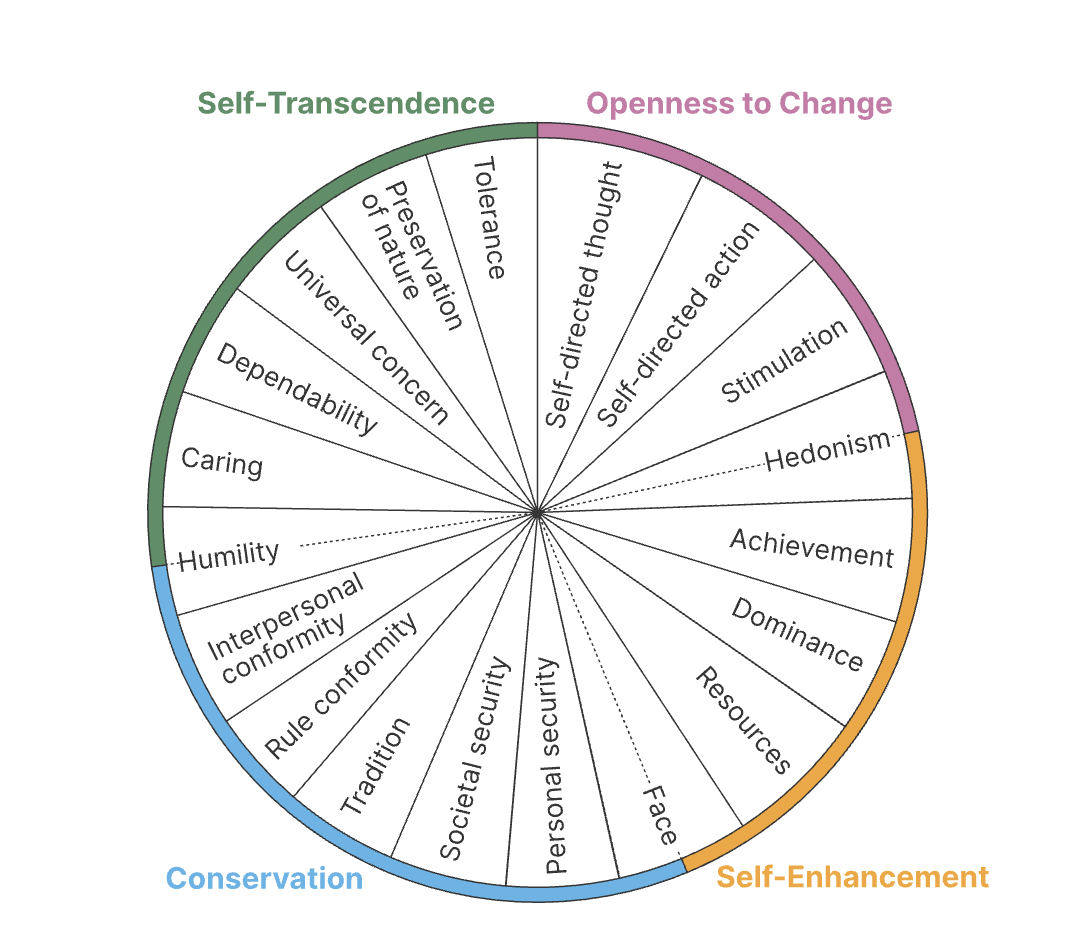}
    \caption{Illustration of Schwartz’s 19 Basic Human Values. Values positioned near one another in the circumplex tend to reflect compatible motivational goals, while those on opposite sides often represent competing motivations. These values are the basis for the 19-dimensional value vector we generate for each comment, based on the ratings produced by the annotation model.}
    \label{fig:value_chart}
\end{figure}

To gauge the expression of values in discourse, we draw on Schwartz's Refined Theory of Basic Human Values \citep{schwartz2012refining}, a well-established framework in cultural psychology. The theory distinguishes 19 conceptually distinct values and organizes them within a circumplex structure that captures systematic compatibilities and tensions among values (See Figure~\ref{fig:value_chart}). For instance, the value Achievement emphasizes individualistic outcomes and personal advancement, whereas the value Universal Concern emphasizes a societal focus. Following prior work~\cite{jahanbakhsh2025value, kolluri2025alexandria}, we identify expressions of human values in CMV conversations using large language models, which enable the analysis of value expression at scale.

In this work, we investigate how value expression and value alignment relate to persuasion in CMV discussions. We ask:

\begin{itemize}
    \item \textbf{RQ1:} Is greater \emph{alignment} between OPs' and commenters' expressed values associated with a higher likelihood of persuasion?
\end{itemize}

An observed association between value alignment and persuasion, however, can arise through different underlying processes. One possibility is \emph{value reframing}. As a conversation unfolds, persuasive commenters may increasingly foreground values that align with those expressed by the OP. Under this account, alignment in expressed values would be weaker or absent in the opening exchange but become more pronounced over the course of a successful interaction.
 
% In this case, we would expect alignment in expressed values to emerge over the course of the conversation, even if such alignment is not apparent in the initial exchange.

Persuasion might alternatively or additionally reflect selection effects, where commenters are more likely to persuade OPs whose values are already similar to their own from the outset. Therefore, we ask:

\begin{itemize}
    \item \textbf{RQ2:} Are the commenters who successfully persuade the OP more likely to engage with OPs whose values are already closer to their own at the outset of the conversation?
\end{itemize}

Finally, even if value alignment is associated with persuasion, we examine whether success in persuasion depends on commenters expressing values that differ substantially from those they typically express in other discussions:

\begin{itemize}
     
    \item \textbf{RQ3:} Does successful persuasion typically involve greater divergence from commenters' typical value expression patterns, compared to unsuccessful persuasion?
\end{itemize}

These distinctions matter because they imply different levers for improving online discourse. If persuasion is largely shaped by who engages with whom, then interventions aimed at fostering constructive disagreement may be most effective at the level of exposure, such as shaping which interactions occur or pairing users with value-compatible interlocutors. If instead persuasion depends on how arguments are framed in response to what the other person cares about, then tools that help participants notice and connect to their interlocutor’s value concerns could support more productive disagreement without requiring agreement on facts or ideology.

\xhdr{Summary of findings}
We find that persuasion on CMV is associated both with pre-existing value compatibility between participants and with value reframing that brings expressed arguments into closer alignment over the course of a conversation. Successful persuasion does not depend on large departures from commenters' typical value expression patterns.

We conclude by offering recommendations for the design of online social spaces.

% By examining how both shared values and evolving value expression relate to persuasion, this study contributes to a deeper understanding of the motivational and framing dynamics that shape online debate. Insights from this work can inform platform design efforts aimed at promoting constructive engagement by highlighting when participants reason from common ground or gradually move toward it. More broadly, understanding how values are expressed, emphasized, and negotiated in conversation offers a pathway toward fostering online discourse that bridges differences rather than deepens divides.

\section{Related Work}

\subsection{Persuasion in Discourse}

Research on persuasion has long sought to understand why some arguments succeed in changing people's views while others fail, across both face-to-face and online conversations~\cite{cialdini2007influence, petty2012communication, fogg2008mass, wachsmuth2017computational}.
Beyond social and contextual influences such as social proof, authority, source credibility, and mood of the listener~\cite{hullett2005impact, mitra2014language, cialdini1999compliance, chaiken2014heuristic}, work on persuasive discourse has examined how properties of arguments and conversational dynamics shape attitude change. This includes the linguistic framing and tone of messages, politeness, the use of evidence and reasoning strategies, and interactional patterns such as number of back-and-forth exchanges, hedging, analytical language and where a contribution appears within an ongoing discussion~\cite{mccroskey1969summary, tan2016winning, ta2022inclusive, althoff2014ask}.

At the same time, these approaches largely characterize persuasion in terms of surface properties of discourse or interactional structure. These features may offer limited insight into why particular arguments resonate with some interlocutors but not others across topics or contexts, or how deeper, more stable dimensions of meaning may shape persuasive dynamics. Relatedly, Perelman’s New Rhetoric in Argumentation Theory posits that persuasion depends on establishing adherence with an audience by appealing to premises, values, and hierarchies of importance that the audience already accepts as legitimate~\cite{long1983role}. From this perspective, arguments succeed not solely because of their logical form or rhetorical polish, but because they invoke considerations that align with the audience's underlying evaluative commitments.

Related work on moral reframing, primarily in the context of political communication, suggests that aligning arguments with an audience's underlying moral principles can increase persuasiveness and the effectiveness of communication~\cite{andrews2017finding, van2015leaders, voelkel2018morally}.
For example, Voelkel and Willer show that economically progressive policy platforms receive greater support from conservative and moderate respondents when framed in terms of values that resonate more strongly with those groups, such as patriotism, family, or tradition~\cite{voelkel2019resolving}. At the same time, prior studies suggest that individuals rarely engage in moral reframing spontaneously even when incentivized with a cash prize to be persuasive~\cite{feinberg2015gulf}, either because they do not consider reframing as a strategy or because they lack the knowledge of how to do so~\cite{feinberg2019moral}.

Our work differs from this line of research in several respects. First, while moral reframing studies typically focus on political issues and pre-defined ideological divides, we examine persuasion across a broad range of topics in naturalistic online discussions. We adopt an individual-level perspective in which values are not assumed to be tightly coupled to partisan identity, and instead treat values as continuous priorities that vary across individuals, even among those who may share similar political affiliations. Finally, rather than studying single-shot persuasive messages, we analyze multi-turn interactions in which value expression is not prescribed in advance but emerges dynamically through conversation.

\subsection{Measuring Expression of Values}

Values transcend specific situations and serve as standards or criteria for what is considered good or bad. 
Schwartz empirically derives a system of values where the trade-offs among broad motivational principles guide attitudes and behaviors across a wide range of contexts. For example, values such as obedience or honesty may be relevant whether one is interacting with friends,
colleagues, or strangers~\cite{schwartz2013value, schwartz1992universals, schwartz2012overview}. This trans-situational quality distinguishes values from norms or attitudes, which tend to be context-specific. Schwartz's system has been applied not only for looking at individuals' values but also for analyzing values in text, including news articles and arguments~\cite{borenstein2025investigating, bardi2008new, kiesel2022identifying}.
This is not the only framework for characterizing values. Others, such as Moral Foundations Theory, offer alternative perspectives~\cite{graham2013moral}. Nonetheless, in this work, we adopt Schwartz because it articulates a broadly encompassing value space in which some values are complementary while others are in systematic tension, which makes it well-suited for studying alignment or misalignment in discourse. Crucially for our setting, this framework treats values as individual-level priorities that vary in relative importance across people, rather than as markers of fixed group membership.

In recent work, there have been efforts to incorporate values into social media ranking algorithms, so that social media feeds can be curated with consideration for certain societal values~\cite{jia2024embedding, bernstein2023embedding} or the values of the end-user~\cite{kolluri2025alexandria, jahanbakhsh2025value}. This requires understanding or labeling the expression of values in social media posts. Prior work has shown that, because value definitions provided by the social sciences are precise constructs, large language models, when prompted with these definitions, can produce value labels that align closely with human annotations~\cite{jia2024embedding, jahanbakhsh2025value, kolluri2025alexandria}. We draw on this work to label the expression of values in the posts and comments on the ChangeMyView subreddit, and examine the relationship of value alignment to persuasion.

\section{Methods}

\subsection{Data}
\label{sec:data}
We use the data from the \emph{Cornell ChangeMyView (CMV) dataset}, a collection of Reddit posts from the ChangeMyView subreddit between Jan 2013 and May 2015~\cite{tan2016winning}.
In each post, an original poster (OP) shares a belief or opinion they hold and invites others to challenge it. Commenters respond with arguments or counterpoints, and if an OP feels that a response has genuinely changed their view, they acknowledge this by awarding the commenter a ``delta'' ($\Delta$) symbol. This explicit feedback provides a rare opportunity to study persuasion in naturalistic, text-based interactions.

Because our work focuses on human values, we filter the dataset down to posts in which values are likely relevant. To do this, we classify each post by broad topic categories including social, ethical, moral, or political issues. We use \texttt{GPT-4o-mini} for this topic classification, as prior work has shown that large language models perform well on text classification and topical labeling tasks~\cite{gilardi2023chatgpt}. This process excludes posts where values are less salient, for instance, discussions about sports. See Appendix for the prompt we used for filtering on the topics that are potentially value-laden.

We focus our analyses on one-on-one conversations between the OP and a single commenter within each thread. For all the analyses, we apply the following inclusion criteria:

\begin{itemize}
    \item \textbf{Active OP engagement:} We include only threads where the last comment in the thread was posted \emph{before} the OP's final comment timestamp across all threads for that post. This ensures that the OP was actively participating at the time the comment was made, and likely saw and considered it.
    \item \textbf{Exactly two participants:} We include only exchanges involving exactly two unique users (the OP and a single commenter) to focus on clearer cases of value alignment and conversational influence without the added complexity of multi-party dynamics.
    
\end{itemize}

For each included thread, we record whether a delta was awarded to the commenter. The delta serves as our primary dependent variable, indicating a successful persuasion event. 

The filtering process results in 10,181 posts and 23,095 users who participate either as OPs or commenters or both. The distribution of deltas received by users in our dataset is as follows: 91.3\% have received 0 deltas, 6.4\% 1 delta, 1.2\% 2 deltas, and 1.1\% 3 or more deltas.

\subsection{Value Annotation}

\paragraph{Value Framework:} 
We use Schwartz’s Refined Theory of Basic Human Values, which defines 19 distinct values, as the conceptual basis for our analysis~\cite{schwartz2012refining}.

\paragraph{Annotation Process:} 
We rate each comment in the dataset on each of the 19 values using a four-level scheme $\{X, 0, 1, 2\}$. We use X for when the value is not expressed. This case denotes absence of evidence (not the lack of importance of the value for the speaker), as some values are simply not invoked by a given topic or an argument. In contrast, ratings 0–2 are used only when the text provides explicit evidence about the (lack of) endorsement of a value. 0 indicates that the comment explicitly opposes or undermines the value, while 1 and 2 indicate weak vs. strong endorsement or expression of that value in the comment.

% Each comment in the dataset is rated for all 19 values. Ratings encode both the \emph{presence} and \emph{salience} of each value using a four-point scale:
% \begin{itemize}
%     \item \textbf{X} = value not expressed
%     \item \textbf{0} = explicitly opposes value
%     \item \textbf{1} = value weakly expressed
%     \item \textbf{2} = value strongly expressed
% \end{itemize}

To preserve conversational context and coherence, we use the \texttt{GPT-4o-mini} model to rate each comment while providing the entire preceding conversation (i.e., the OP post and the resulting back and forth between a commenter and the OP in a thread) as input. The prompt for labeling each comment for values is given in the Appendix.

\paragraph{Value Vectors:} 

We represent each user's value expression in a thread as a 19-dimensional \emph{value vector}, where each dimension corresponds to one of the Schwartz values. We initialize a vector from a user's first contribution in the thread (the post for OPs, the first comment for commenters). We annotate each subsequent comment \emph{sequentially}, with the full preceding conversation provided as context to the annotation model. We retain only a single aggregated value vector per user per thread.

Let $r_{u,t}^{(k)} \in \{X, 0, 1, 2\}$ denote the rating assigned to value $k$ for user $u$’s $t$-th comment, produced given the preceding conversation history.  For each value dimension, the update rule is:

\[
v_{u}^{(k)} \leftarrow
\begin{cases}
v_{u}^{(k)} & \text{if } r_{u,t}^{(k)} = X \\
0 & \text{if } r_{u,t}^{(k)} = 0 \\
r_{u,t}^{(k)} & \text{if } v_{u}^{(k)} = X \\
\max\!\left(v_{u}^{(k)}, r_{u,t}^{(k)}\right) & \text{otherwise}
\end{cases}
\]

We update the user’s value vector incrementally as comments are processed, but do not store intermediate vectors. This means that the final aggregated vector $\mathbf{v}_u$ is retained for analysis. The intuition behind the update rules is that strong endorsements persist unless explicitly rejected, and values not mentioned does not alter prior evidence. Each OP has one value vector per thread they participate in (i.e., one per commenter they respond to). Commenters typically engage in only one thread per post, so they usually have a single value vector per post. Out of the 18,217 unique commenters, only 1\% participated in more than one thread under the same post.

\paragraph{Binarization.} For the purposes of analysis, we collapse weak and strong expressions of a value (ratings of 1 and 2) into a single binary indicator of presence, while preserving the \texttt{X} (not expressed) and \texttt{0} (explicitly opposed) labels. This decision is guided by two considerations. First, prior work adopts a similar procedure where values are labeled on a ternary scale of not expressed, weakly expressed, and strongly expressed, and these labels are subsequently binarized to evaluate agreement with human annotations. Under this binarization, large language model annotations show high agreement with human judgments \citep{kolluri2025alexandria}. Second, recognizing and interpreting values in social media posts involves an inherent degree of subjectivity, particularly with respect to fine-grained distinctions in salience \citep{epstein2025measuring}.
In our context, where we are interested in how participants might respond to each other's expressed values, binary presence may offer a more reliable signal than subtle gradations of strength. We therefore binarize post-hoc to reduce noise.

\subsection{Analytic Approach}

All analyses for RQ1--3 were pre-registered.\footnote{Pre-registration link: \url{https://osf.io/9kz6n/overview?view_only=7a9e3acd162f4846ada4ba89906477b0}} Sample sizes were determined a priori according to the pre-registration plan. We z-scored all value distance measures prior to modeling to improve interpretability. Coefficients therefore reflect effects per standard deviation increase.

We create three separate datasets to answer the questions: one for RQ1, one for both RQ2 and RQ3, and one consisting of high and low delta earners. We describe the procedure for each in their corresponding sections below. Table~\ref{tab:descriptive_stats} reports the descriptive statistics for each dataset, including the average number of comments per thread and the average number of active values expressed in thread-level vectors, where active values are defined as values that receive non-\texttt{X} ratings. Delta-awarded threads tend to be longer and exhibit more expressed values. We do not include thread length as a control in our models, as it is plausibly endogenous to persuasive dynamics and may constitute part of the mechanism through which value alignment emerges over interaction.

\subsubsection{RQ1: Emergent Value Alignment and Persuasion.}
For the RQ1 analyses, we filter the dataset down to posts for which there exists at least one thread that was awarded a delta and one thread that was not.
We sample 135 posts from this dataset, along with all their associated threads that meet our inclusion criteria.
To examine whether value alignment is associated with persuasion and whether such alignment emerges through interaction, we measure \emph{value misalignment} at two points in the conversation. In both cases, we operationalize misalignment as the mean absolute error (MAE) between the OP’s and the commenter’s value vectors, computed over value dimensions where both users express (or explicitly oppose) a value. We therefore operationalize alignment based on overlap in explicitly expressed values, and consider non-mention of a value as uninformative rather than as misalignment.
Lower MAE indicates greater alignment.

 We compute \textsc{Initial Misalignment} based on the OP’s post and the commenter’s first reply. This captures the degree of value alignment present on the very first exchange. We compute \textsc{Cumulative Misalignment} based on each participant’s value vector after the full conversation. These vectors reflect the values expressed across all comments in the thread and capture alignment as it stands at the end of the interaction.

Comparing initial and cumulative misalignment allows us to examine whether alignment in expressed values increases over the course of an interaction. Alignment that is weaker or absent in the opening exchange but stronger after the full conversation is consistent with \emph{value reframing},  which we interpret as increased emphasis on values that resonate with the interlocutor as the conversation unfolds. To test these relationships, we fit separate logistic mixed-effects models predicting whether a delta was awarded, with random intercepts for the post and the commenter.

\begin{table*}[t]
\centering
\caption{Descriptive statistics for datasets used in each of the analyses. Active values are defined as values receiving non-\texttt{X} ratings. }
\label{tab:descriptive_stats}
{\renewcommand{\arraystretch}{1.25}
\begin{tabular}{llrrrr}
\hline
\textbf{Dataset} & \textbf{Delta Awarded} & \textbf{Avg thread length} & \multicolumn{2}{c}{\textbf{Avg \# of Active values}} \\
\cline{4-5}
& & &  \textbf{Commenter} & \textbf{OP} \\
\hline
RQ1 
& Yes & 3.78 (1.56) &  6.26 (2.62) & 8.37 (2.51) \\
& No  & 2.58 (1.35) & 5.09 (2.75) & 7.30 (2.80) \\
\hline
RQ2 \& RQ3
& Yes &  3.21 (1.80)  &  2.36 (1.71)\\
& No  &  2.09 (1.95) &  1.97 (1.64)\\
\hline
High vs.\ Low Delta Earners subset
& Yes &  3.28 (1.85)  & 2.28 (1.69)\\
& No  &  2.12 (2.00) &  1.99 (1.69)\\
\hline
\end{tabular}
}
% \vspace{0.5em}
% \small
% \textit{Note.} 
\end{table*}

\subsubsection{RQ2: Pre-existing Value Alignment}

To investigate RQ2 and RQ3, we construct a sample of 704 unique commenters stratified by how many deltas they have received across the dataset: 0, 1, 2, or 3+. Without this stratification, the sample would be dominated by users who rarely succeed.

From each group, we sample up to a target cap of eligible users to obtain a roughly balanced distribution across delta groups. Eligibility requires participation in at least three threads across three different posts. If a delta group contains fewer eligible users than the cap, we reallocate the remaining quota to other delta groups. The same sample is used for the analyses related to RQ2 and RQ3. This set of commenters had contributed 11967 threads overall (median of 7 per person).

\begin{figure}[t]
    \centering
    \includegraphics[width=\linewidth]{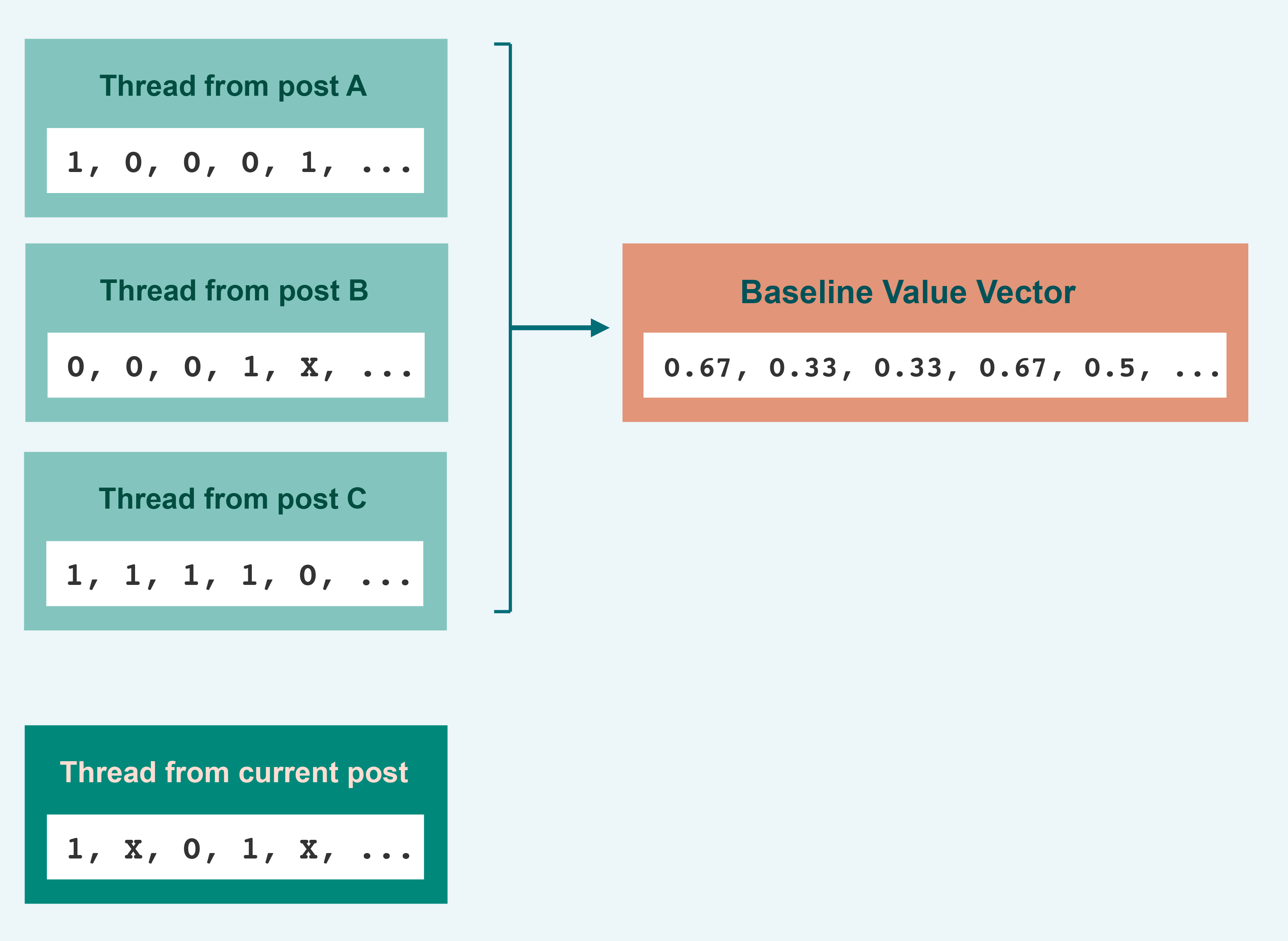}
    \caption{Visualization of how a Baseline Value Vector for a commenter is constructed.}
    \label{fig:baseline_vector}
\end{figure}

The threshold of participation in at least three threads is so that we can get a more stable estimate of a user's \textsc{baseline value vector}, which we define as the average of their value vectors from all threads in other posts other than the one being analyzed (see Figure~\ref{fig:baseline_vector}). We do not interpret this baseline as the commenter’s intrinsic value profile. Rather, it serves as a reference point for the values a commenter tends to express outside the local context of the focal conversation. Aggregating across other threads provides an estimate of the commenter's typical value expression, against which we interpret within-thread shifts.

By requiring at least three threads, we guarantee that each baseline is computed from at least two other observations. When computing this average, we ignore dimensions marked as \texttt{X} (not expressed) in a given value vector for the user. Therefore, for each value dimension, we average only over the value vectors in which that value was explicitly expressed (as either 0 or 1).

To test whether successful persuasion is more likely when commenters engage with OPs whose values are already similar to their own, we use \textsc{Baseline Value Distance} as the predictor. This variable captures the mean absolute error (MAE) between the OP's initial value vector (from their post) and the commenter's Baseline Value Vector. Lower Baseline Value Distance indicates greater pre-existing value alignment between the commenter and the OP before the conversation begins.

We fit a logistic mixed-effect model to predict if a delta is awarded. We include random intercepts for the post and the commenter to account for repeated measures.

\subsubsection{RQ3: Divergence from Typical Value Expressions.}

RQ3 examines the extent to which commenters depart from their own baseline value expressions when participating in a given thread.

To operationalize this, we compute \textsc{Normalized Shift}, which we define as the mean absolute error (MAE) between a commenter's in-thread value vector and their Baseline Value Vector, normalized by the Baseline Value Distance (i.e., the MAE between the commenter's Baseline Value Vector and the OP's initial post). Intuitively, this measure captures how much a commenter's expressed values in a given conversation deviate from their typical expression patterns, relative to how different they were from the OP at the outset. This normalization accounts for differences in how misaligned a commenter is with the OP at the outset of the conversation. Commenters who begin far from the OP have greater potential to change their expressed values than those who begin relatively aligned. By scaling in-thread divergence by this initial distance, Normalized Shift enables comparisons across conversations that start at different levels of value misalignment.

Higher values of Normalized Shift indicate greater divergence from a commenter's baseline value expression, while lower values indicate that the commenter's in-thread value expression remains closer to what they typically express across other discussions.

We fit a logistic mixed-effects model with whether delta is awarded as the binary outcome, and include random intercepts for the post and the commenter.

\subsubsection{RQ2 and RQ3 Effects for High- and Low-Delta Earners.}

We conduct user-level analyses to examine whether patterns of Baseline Value (Mis)Alignment and value shift differ between commenters who are more versus less successful at earning deltas. We test whether Baseline Value Distance and Normalized Shift, computed across multiple threads per commenter, distinguish commenters who are generally more or less successful at persuasion.

We contrast two groups of commenters: \emph{low delta earners}, sampled from users who receive zero deltas in the dataset, and \emph{high delta earners}, sampled from users who receive three or more deltas. The sample for each group contains 226 users, who, in total, have contributed 10825 threads (median of 9 per person). We pre-registered 276 high- and 276 low-delta earners; but after applying inclusion criteria, only 226 high-delta earners remained eligible, so we used all 226 and sampled 226 low-delta earners.

We focus on these extreme groups because the dataset represents a finite snapshot of CMV activity over approximately 2.5 years, and users with intermediate delta counts may be misclassified due to activity occurring outside the observed window. We then test whether commenters’ average Baseline Value Distance and average Normalized Shift, aggregated across their threads, predict membership in these groups. While this question was pre-registered, we revised the analysis approach to aggregate predictors across threads for each commenter, rather than modeling individual threads, because the outcome (high- versus low-delta earner status) is defined at the commenter level. For each predictor, we fit a logistic regression model predicting high- versus low-delta earner status for a user, including the number of threads the user participated in as a control variable.

\section{Results}

% \begin{itemize}
%     \item \textbf{RQ1:} Threads contained an average of 1.81 comments per thread overall (SD = 1.47). Delta-awarded threads averaged 2.78 comments (SD = 1.56), while non-delta-awarded threads averaged 1.58 comments (SD = 1.35). In delta-awarded threads, commenters expressed a mean of 6.26 values (SD = 2.62), compared to 5.09 values in non-delta threads (SD = 2.75). OPs expressed a mean of 8.37 values in delta-awarded threads (SD = 2.51), compared to a mean of 7.30 values in non-delta awarded threads (SD = 2.80).
    
%     \item \textbf{RQ2 and RQ3:} Threads contained an average of 2.23 comments (SD = 1.97) and 2.02 active values per thread overall (SD = 1.65). Delta-awarded threads averaged 3.21 comments (SD = 1.80) and 2.36 active values (SD = 1.71), while non-delta-awarded threads averaged 2.09 comments (SD = 1.95) and 1.97 active values (SD = 1.64).
    
%     \item \textbf{RQ2 and RQ3 Effects for High- and Low-Delta
% Earners:} Threads contained an average of 2.27 comments (SD = 2.02) and 2.02 active values per thread overall (SD = 1.69). Delta-awarded threads averaged 3.28 comments (SD = 1.85) and 2.28 active values (SD = 1.69), while non-delta-awarded threads averaged 2.12 comments (SD = 2.00) and 1.99 active values (SD = 1.69).
% \end{itemize}

\subsection{RQ1: Emergent Value Alignment and Its Relationship with Persuasion}

We find that Cumulative Value Misalignment between the OP and the commenter is a significant predictor of successful persuasion. Recall that Cumulative Value Misalignment is computed from the value vectors after the full exchange, reflecting all values expressed over the course of the conversation. Greater Cumulative Misalignment is associated with a lower likelihood of receiving a delta ($\beta= -0.38$, $SE = 0.11$, $z = -3.55$, $p < 0.001$).

\begin{figure}[t]
    \centering
    \includegraphics[width=\linewidth]{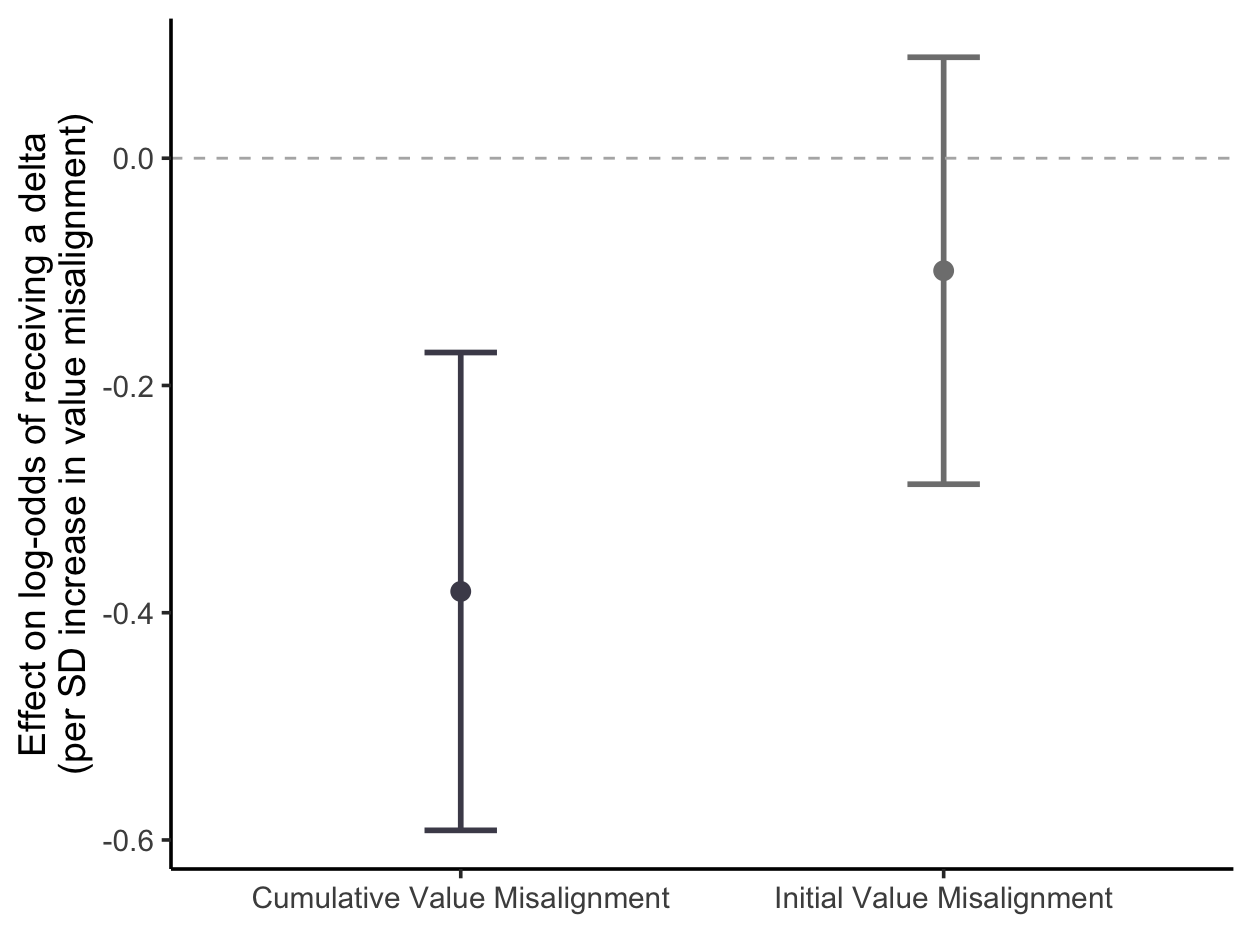}
    \caption{Effects of value misalignment on the likelihood of persuasion. Points show standardized coefficients from logistic mixed-effects models predicting whether a comment receives a delta; error bars indicate 95\% confidence intervals. Initial Value Misalignment, computed from the OP's post and the commenter's first reply, is not significantly associated with persuasion. On the other hand, Cumulative Value Misalignment over the full conversation is a significant negative predictor. Coefficients reflect the effect of a one–standard deviation increase in value misalignment.}
    \label{fig:RQ1}
\end{figure}

By contrast, Initial Value Misalignment, which we compute from the OP's post and the commenter's first reply, is not significantly associated with successful persuasion ($\beta= -0.10$, $SE = 0.10$, $z = -1.03$, $p=0.30$). See Figure~\ref{fig:RQ1}.

These patterns suggest that value alignment associated with successful persuasion is not apparent in the opening exchange but emerges over the course of the interaction, consistent with \emph{value reframing} as we operationalize it in this work.

\subsection{RQ2: The Effect of Pre-existing Value Alignment on Persuasion}

\begin{figure}[t]
    \centering
    \includegraphics[width=\linewidth]{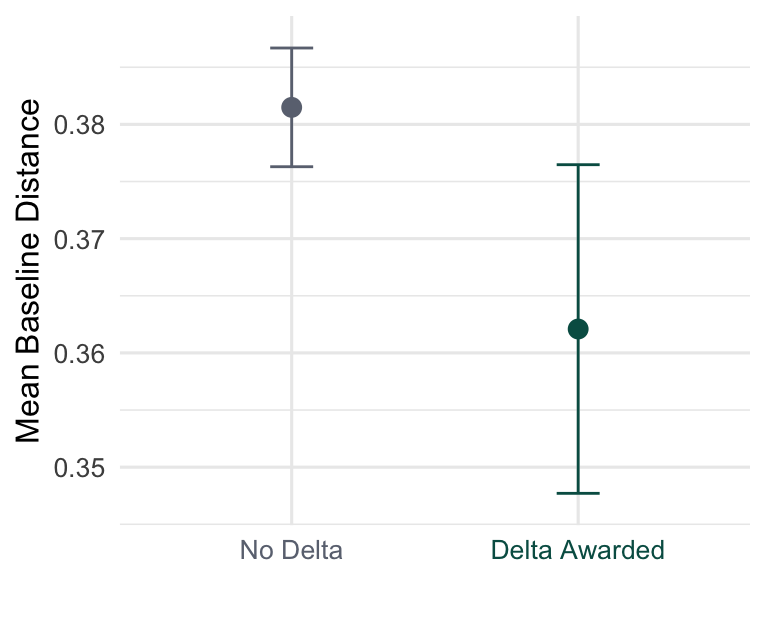}
    \caption{Mean pre-existing value distance between commenters and original posters for comments that did and did not receive a delta. Error bars indicate 95\% confidence intervals. Comments that receive a delta tend to exhibit greater pre-existing value compatibility between participants.}
    \label{fig:RQ2}
\end{figure}

Baseline Value Distance captures the distance between the OP's initial value expression and the commenter's typical value expression across other threads, reflecting pre-existing value alignment prior to the conversation. We find that Baseline Value Distance is a significant negative predictor of successful persuasion ($\beta = -0.08$, $SE = 0.03$, $z = -2.31$, $p = 0.021$), indicating that commenters are more likely to receive a delta when their values are already closer to those of the OP at the outset. See Figure~\ref{fig:RQ2}.

This result suggests that persuasion on CMV is partly shaped by selection effects: successful persuasion is more likely to occur when commenters engage with OPs whose value priorities are already relatively compatible with their own, rather than when commenters and OPs begin from substantially different value positions.

\subsection{RQ3: The Effect of Divergence from Typical Value Expression on Persuasion}

\begin{figure}[h]
    \centering
    \includegraphics[width=\linewidth]{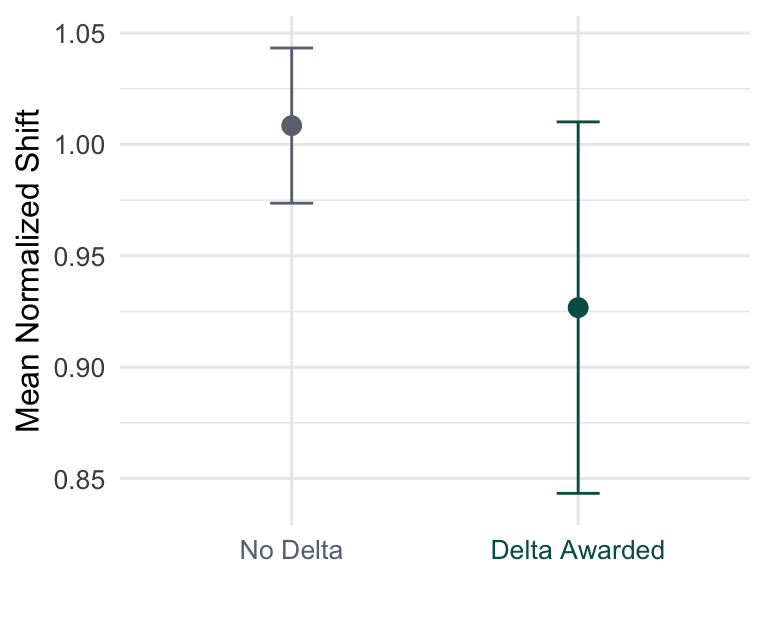}
    \caption{Mean normalized shift in commenters’ expressed values for comments that did and did not receive a delta. Error bars indicate 95\% confidence intervals. Successful persuasion is not associated with greater divergence from commenters' typical value expression patterns.}
    \label{fig:RQ3}
\end{figure}

We find no statistically significant evidence that greater divergence from a commenter's typical value expression patterns is associated with successful persuasion. Normalized Shift is not a significant predictor of whether a comment receives a delta ($\beta = -0.06$, $SE = 0.05$, $z = -1.41$, $p = 0.16$), indicating that successful commenters do not systematically depart from their usual value expression patterns. See Figure~\ref{fig:RQ3}.

\subsection{RQ2 and RQ3 Effects for High- and Low-Delta Earners}

At the user level, we find no evidence that average patterns of value alignment or value shifting distinguish commenters who are generally more successful at earning deltas from those who are not. Mean Baseline Value Distance, averaged across each commenter's threads, is not significantly associated with high- versus low-delta earner status ($\beta = -0.03$, $SE = 0.14$, $z = -0.21$, $p = 0.84$). Similarly, mean Normalized Shift is not a significant predictor of delta earner status ($\beta = -0.14$, $SE = 0.14$, $z = -0.98$, $p = 0.33$).

In both models, the number of threads a commenter participates in is a strong positive predictor of high-delta earner status ($\beta = 0.26$, $SE = 0.03$, $p < 0.001$ and $\beta = 0.25$, $SE = 0.03$, $p < 0.001$, respectively), suggesting greater opportunity to receive deltas.

\section{Discussion}

Our findings suggest that value-based alignment shapes persuasion in online discussion in two complementary ways. First, persuasion is more likely when commenters and OPs are broadly compatible in their value priorities, even before the conversation begins (\textbf{RQ2}). At the same time, successful persuasion is associated with lower Cumulative Value Misalignment over the course of a conversation, despite the absence of value alignment in the initial exchange (\textbf{RQ1}). This means that the alignment in expressed values that emerges through the back-and-forth of the conversation is characteristic of successful persuasion. These findings point to a process in which commenters increasingly foreground value considerations that align with those expressed by the OP as the interaction unfolds. This asymmetry reflects the structure of CMV discussions, where commenters are positioned as persuaders and OPs as evaluators of successful persuasion.

Importantly, this form of value framing differs from how the concept has typically been studied in prior work. Much of the literature on moral or value reframing focuses on deliberate appeals tailored to broad ideological or partisan groups, often in one-shot persuasive messages~\cite{van2015leaders, voelkel2018morally, feinberg2019moral}. By contrast, our results indicate that value reframing can emerge at a finer granularity, shaped by the dynamics of a specific interaction and grounded in individual-level value priorities rather than group identity.

One possible interpretation of these findings draws on Communication Accommodation Theory, which posits that speakers often adapt aspects of their communicative behavior such as language use, framing, or emphasis, to converge toward their interlocutor in order to facilitate understanding, affiliation, or approval~\cite{gallois2005communication, giles2013communication}. From this perspective, shifts in expressed values may reflect a form of accommodative behavior, where commenters adjust which underlying priorities they foreground in response to the values expressed by the original poster. Crucially, such adjustment need not involve large departures from a commenter's typical value profile. Conditional on the Baseline Value Distance between participants, greater divergence from one's own baseline value expression is not associated with a higher likelihood of persuasion (\textbf{RQ3}).

Our results do not imply that commenters strategically or consciously reframe their arguments to match the OP's values, similar to how Accommodation is not necessarily explicitly planned. Rather, value accommodation may emerge implicitly through interaction, as participants attend to what appears to matter to their interlocutor and emphasize compatible value considerations in response. Such accommodation could increase the likelihood that arguments are perceived as relevant or legitimate, even when partial disagreement remains.

One apparent tension in our findings is that Baseline Value Distance predicts persuasion, whereas value alignment in the first exchange does not. We believe this reflects a difference between value compatibility and when such compatibility becomes visible in discourse. Values often function as underlying premises rather than opening moves in argumentation, and early turns in a conversation may focus on clarifying claims, probing assumptions, or establishing conversational footing rather than explicitly invoking shared priorities. As a result, values that are compatible between participants may not be foregrounded in the first exchange, even when they later play a role in shaping persuasive outcomes. This helps explain why value compatibility between interlocutors predicts persuasion despite not being apparent at the outset of the conversation.

\subsection{Design Implications}

A growing body of work suggests that polarization on social media is not driven solely by insulating users from opposing views, but also by exposure to countervailing content that feels misaligned with or adversarial to users' underlying concerns~\cite{tornberg2022tweeting, bail2022breaking}. At the same time, limiting exposure to counterattitudinal content raises concerns about filter bubbles and echo chambers, in which individuals' existing views go largely unchallenged and are reinforced through repeated affirmation. Our findings suggest that framing arguments in the values of an interlocutor can help them engage more constructively with counterarguments and recognize their merits, even when they initially oppose the position being advanced. This points to a potential role for social media ranking algorithms to, amid a wide range of opposing content, preferentially surface content that are already grounded in values that resonate with the user. Highlighting such content may support engagement across disagreement lines without exacerbating polarization.

Our work suggests that value alignment and consequently, acceptance of counter-arguments can emerge through interaction. We call on future work to explore how online social platforms might help participants recognize and support this process.
One possible direction is to explore tools that assist with \emph{value translation} during an ongoing exchange, by helping users re-express their arguments in ways that connect to the values their interlocutor has already made visible in the conversation. An important challenge for such a mechanism is ensuring that value translation functions as a reflective aid for both parties, rather than being perceived as a strategic device for persuasion or manipulation.

\subsection{Limitations and Future Work}

We highlight limitations of our study that can also inform future work in this
area. First, our analyses are observational and rely on the explicit awarding of deltas as a proxy for persuasion. While deltas provide a rare and valuable signal of attitude change, they capture only a subset of persuasive outcomes and may miss more partial or unacknowledged shifts in perspective. Second, our operationalization of values relies on large language models to infer value expression from text. Although prior work shows strong agreement between such annotations and human judgments~\cite{jahanbakhsh2025value, kolluri2025alexandria}, value expression in discourse remains inherently subjective~\cite{epstein2025measuring}, and our measures should be understood as approximations rather than direct observations of participants' internal value priorities. Future work could further validate these measures by combining them with participant-reported value priorities. Finally, our focus on one-on-one exchanges in r/ChangeMyView enables clearer attribution of influence but limits generalizability to more complex, multi-party discussions or to platforms with different interaction norms. Future work should explore value reframing in multi-party discussions, real-time conversations, or feed-based platforms where users encounter and react to content outside of a shared conversational thread.

\section{Ethical Considerations}

Our findings highlight value alignment or reframing as a mechanism associated with successful persuasion, but persuasion itself is not always a desirable or benign outcome. Techniques that increase the effectiveness of persuasive communication can also be used in manipulative ways, particularly in contexts involving political messaging, advertising, or where there is power imbalance at play. Framing arguments around an individual's values may lower resistance to counterarguments, but it may also bypass critical reflection or be used to exploit vulnerabilities rather than support mutual understanding. As such, value alignment should be treated as a mechanism whose ethical implications depend on how and why it is deployed.

Looking ahead, if future systems were to incorporate value inference or value-aware curation inspired by findings such as ours, additional ethical concerns would arise. Automated or inferred representations of users' values may be inaccurate, overly reductive, or perceived as intrusive, especially given the personal and identity-linked nature of values. Moreover, prioritizing value-aligned content risks reinforcing value silos or advantaging users who are already more articulate or whose values are more socially legible. These considerations underscore the importance of designing value-aware systems that support transparency, user agency, and reflection, rather than optimizing persuasion or engagement alone.

\bibliography{aaai2026}

\subsection{Paper Checklist}

\begin{enumerate}

\item For most authors...
\begin{enumerate}
    \item  Would answering this research question advance science without violating social contracts, such as violating privacy norms, perpetuating unfair profiling, exacerbating the socio-economic divide, or implying disrespect to societies or cultures?
    \answerYes{Yes.}
  \item Do your main claims in the abstract and introduction accurately reflect the paper's contributions and scope?
    \answerYes{Yes.}
   \item Do you clarify how the proposed methodological approach is appropriate for the claims made? 
    \answerYes{Yes. See the Introduction and the Methods.}
   \item Do you clarify what are possible artifacts in the data used, given population-specific distributions?
    \answerNA{NA}
  \item Did you describe the limitations of your work?
    \answerYes{See Sec. Limitations.}
  \item Did you discuss any potential negative societal impacts of your work?
    \answerYes{Yes, see Sec. Ethical Considerations.}
      \item Did you discuss any potential misuse of your work?
    \answerYes{Yes, see Sec. Ethical Considerations.}
    \item Did you describe steps taken to prevent or mitigate potential negative outcomes of the research, such as data and model documentation, data anonymization, responsible release, access control, and the reproducibility of findings?
    \answerYes{Yes. We discuss potential risks and societal impacts of value-based persuasion in the Ethical Considerations section. We rely on an existing, anonymized dataset and do not release new models or user-level value profiles. We report the LLM prompts used for content labeling in the Appendix.}
  \item Have you read the ethics review guidelines and ensured that your paper conforms to them?
    \answerYes{Yes.}
\end{enumerate}

\item Additionally, if your study involves hypotheses testing...
\begin{enumerate}
  \item Did you clearly state the assumptions underlying all theoretical results?
    \answerYes{Yes. We state the assumptions underlying our empirical analyses (e.g., operationalization of persuasion, value inference, and alignment measures) in the Methods section.}
  \item Have you provided justifications for all theoretical results?
    \answerYes{Yes. We justify our empirical results through statistical analysis and theoretically grounded interpretation in the Discussion.}
  \item Did you discuss competing hypotheses or theories that might challenge or complement your theoretical results?
    \answerYes{Yes. We explicitly discuss alternative mechanisms such as selection effects versus emergent value reframing (RQ1 vs. RQ2) in both the Introduction and Discussion.}
  \item Have you considered alternative mechanisms or explanations that might account for the same outcomes observed in your study?
    \answerYes{Yes. We consider selection effects, conversational dynamics, and accommodation-based explanations in the Discussion section.}
  \item Did you address potential biases or limitations in your theoretical framework?
    \answerYes{Yes. We discuss limitations related to observational data, value inference, and generalizability in the Limitations section.}
  \item Have you related your theoretical results to the existing literature in social science?
    \answerYes{Yes. We situate our findings in relation to work on persuasion, moral reframing, and Communication Accommodation Theory in the Related Work and Discussion.}
  \item Did you discuss the implications of your theoretical results for policy, practice, or further research in the social science domain?
    \answerYes{Yes. We discuss implications for platform design, content curation, and future research directions in the Design Implications and Limitations \& Future Work sections.}
\end{enumerate}

\item Additionally, if you are including theoretical proofs...
\begin{enumerate}
  \item Did you state the full set of assumptions of all theoretical results?
    \answerNA{NA}
	\item Did you include complete proofs of all theoretical results?
    \answerNA{NA}
\end{enumerate}

\item Additionally, if you ran machine learning experiments...
\begin{enumerate}
  \item Did you include the code, data, and instructions needed to reproduce the main experimental results (either in the supplemental material or as a URL)?
    \answerYes{Yes, we use a publicly available dataset and describe all preprocessing and analysis steps in detail. See Appendix for model prompts.}
  \item Did you specify all the training details (e.g., data splits, hyperparameters, how they were chosen)?
    \answerNA{We do not train our own machine learning model. We use \texttt{GPT-4o-mini}.}
     \item Did you report error bars (e.g., with respect to the random seed after running experiments multiple times)?
    \answerYes{Yes.}
	\item Did you include the total amount of compute and the type of resources used (e.g., type of GPUs, internal cluster, or cloud provider)?
    \answerYes{Yes, when describing the system, we discuss that we use \texttt{GPT-4o-mini} via the OpenAI API.}
     \item Do you justify how the proposed evaluation is sufficient and appropriate to the claims made? 
    \answerYes{Yes, our evaluation directly aligns with our research questions.}
     \item Do you discuss what is ``the cost`` of misclassification and fault (in)tolerance?
    \answerYes{Yes. We discuss the implications of value misclassification and subjectivity in the Limitations and Ethical Considerations sections.}
  
\end{enumerate}

\item Additionally, if you are using existing assets (e.g., code, data, models) or curating/releasing new assets, \textbf{without compromising anonymity}...
\begin{enumerate}
  \item If your work uses existing assets, did you cite the creators?
    \answerYes{Yes. We cite the creators of the CMV dataset and prior work used for value annotation.}
  \item Did you mention the license of the assets?
    \answerYes{Yes. The CMV dataset is publicly available and used in accordance with its stated terms.}
  \item Did you include any new assets in the supplemental material or as a URL?
    \answerNA{NA}
  \item Did you discuss whether and how consent was obtained from people whose data you're using/curating?
    \answerNA{NA}
  \item Did you discuss whether the data you are using/curating contains personally identifiable information or offensive content?
    \answerNA{NA}
\item If you are curating or releasing new datasets, did you discuss how you intend to make your datasets FAIR?
\answerNA{NA}
\item If you are curating or releasing new datasets, did you create a Datasheet for the Dataset? 
\answerNA{NA}
\end{enumerate}

\item Additionally, if you used crowdsourcing or conducted research with human subjects, \textbf{without compromising anonymity}...
\begin{enumerate}
  \item Did you include the full text of instructions given to participants and screenshots?
    \answerNA{Not applicable. This study does not involve direct interaction with human participants.}
  \item Did you describe any potential participant risks, with mentions of Institutional Review Board (IRB) approvals?
    \answerNA{Not applicable. We analyze publicly available, anonymized data and did not recruit participants. Our study was deemed exempt by our IRB.}
  \item Did you include the estimated hourly wage paid to participants and the total amount spent on participant compensation?
    \answerNA{NA}
   \item Did you discuss how data is stored, shared, and deidentified?
   \answerNA{We rely on anonymized, publicly released data and do not store or release additional identifying information.}
\end{enumerate}

\end{enumerate}

\clearpage

\appendix
\renewcommand{\thelstlisting}{A.\arabic{lstlisting}}
\setcounter{lstlisting}{0}  % Reset listing counter

\section{Appendix}
% \begin{lstlisting}[style=prompt, caption={hi}, label={lst:prompt}]

% \end{lstlisting}

\begin{tcolorbox}[breakable, colback=teal!10, colframe=teal!80, title= Prompt for determining if the post topic is relevant,   before upper={
    \small\emph{
      The prompt we used to label whether a post topic is value-laden and therefore relevant to our analyses.
    }
    \medskip\hrule\medskip
  }]
\scriptsize

   Please analyze the following Reddit post and determine if it discusses social, ethical, moral, or political issues.
    A topic should be considered relevant if it involves:
    \begin{itemize}
        \item Social issues (discrimination, inequality, community, culture, social justice)
        \item Ethical/moral questions (right vs wrong, justice, fairness, responsibility, values)
        \item Political issues (government, policy, law, rights, democracy, authority)
        \item Current social topics (climate, immigration, healthcare, education, gender, religion)
        \item Behavioral/psychological topics with social implications
        \item  Economic issues with social/political implications
    \end{itemize}
 
    Examples of RELEVANT topics:
    \begin{itemize}
        \item Discussions about discrimination or inequality
        \item Questions about moral behavior or ethics
        \item Political opinions or government policy
        \item Social justice issues
        \item Religious or cultural conflicts
        \item Environmental or climate discussions
        \item Healthcare or education policy
        \item Economic inequality or workplace issues
    \end{itemize}
    
    Examples of NOT RELEVANT topics:
    \begin{itemize}
        \item Technical questions about programming, math, or science
        \item Personal relationship advice without broader social implications
        \item Entertainment, movies, games, or hobbies
        \item Simple factual questions
        \item Product recommendations or reviews
    \end{itemize}
    
    Post to analyze:
    \verb|{POST}|
    
    Please respond with only ``YES'' if this topic is relevant to social, ethical, moral, or political issues, or ``NO'' if it is not relevant.

\end{tcolorbox}

\begin{tcolorbox}[breakable, colback=teal!10, colframe=teal!80, title= Prompt for labeling the values of a comment,
before upper={
    \small\emph{
      The prompt we used to label the values expressed in a comment, given the preceding conversation history.
      We replace \texttt{conceptDefinitions} with a dictionary of value–definition pairs. We took the value names and definitions from~\cite{jahanbakhsh2025value} which lightly adapts Schwartz's original definitions through consultation with a cultural psychologist to improve clarity and interpretability.
      Each \texttt{example} variable is assigned its corresponding comment and conversation history.
    }
    \medskip\hrule\medskip}]

\vspace{0.5em}

\scriptsize

    Consider the following set of concepts, listed as CONCEPT : DEFINITION
    
    \verb|{conceptDefinitions}|
        
    You are given a Reddit comment and the entire preceeding conversation if any. Your task is to evaluate single Reddit comment based on the conversation context preceeding it if any.
    For each concept, rate how the comment reflects the concept on a Likert scale with options [0, 1, 2, ``X''], where:

    \begin{itemize}
        \item 0 = This comment explicitly goes against or undermines this concept
        \item 1 = This comment reflects this concept a little bit
        \item 2 = This comment reflects this concept strongly
        \item ``X'' = This comment makes **no reference at all** to this concept (neutral absence, not rejection)
    \end{itemize}
      
    Your goal is to evaluate how the comment interacts with and reflects value themes present in the conversation.
    Use the preceding value signals to guide your rating --- especially if the comment builds on, agrees with, reframes, or pushes back against them.
    
    IMPORTANT TIPS
    \begin{itemize}
        \item Do not treat it in isolation — consider the entire prior conversation thread to understand:
        Which values were made salient or contested?
        How the final comment responds to, builds on, or diverges from earlier value expressions?
        
        \item Your ratings must be based solely on the actual **content and opinions expressed by the user in the comment**, not on their conversational style, tone, or interaction behavior.

        \item Do NOT assign ratings based on the fact that the user is asking questions, inviting discussion, seeking clarification, or responding politely. These are interaction dynamics, not expressions of values.

        \item Focus strictly on whether the user’s comment promotes, describes, or defends the concept itself.

        \item For example, do not award a rating for  “Independent thoughts” or “Independent actions” simply because the user questions or challenges an idea or belief. Only rate these if the user explicitly promotes independent thinking or decision-making within society.

        \item Also, do not merely give a numerical rating to a value just because it is present in the conversation. Analyze how the comment actually expresses the value. For example, if the value “Wealth” is expressed, give a rating based on if the user is against or for it.
          
    \end{itemize}
       
    Output: 
    
    Each individual comment should have a JSON dictionary of the following format. Rating is a dictionary of key value pairs, with each key being a concept and each value your rating for that concept and the reasoning behind that comment. Only output the JSON dictionary in your final output.
    
    \verb|{ ``Rating'': { ``Concept'': ``RATING'' } }|
    
\vspace{1em}
    Here are two specific JSON dictionaries for a singular comment:

\vspace{1em}

\verb|{example1}|

\begin{lstlisting}[  basicstyle=\ttfamily\scriptsize,
  breaklines=true,
  breakindent=0pt,
  breakautoindent=false,
  xleftmargin=0pt,
  numbers=none,
  columns=fullflexible] 
{
 ``Rating'': {
    ``Reputation'': ``X because this post doesn't focus on anyone being public approval or on protecting public image'', 
    ``Power'': ``0 because the post argues for the reduction of dominance and authoritarian influence, particularly in institutions like the police and schools'', 
    ``Wealth'': ``X because the post has no correlation to obsession about gaining money or material status'', 
    ``Achievement'': ``X because the post doesn't focus on ambition, success, or striving for accomplishment'', 
    ``Pleasure'': ``X because the post doesn't reference enjoyment, fun, or hedonistic values'', 
    ``Independent thoughts'': ``2 because the post focuses heavily on wanting students to form their own opinions about America and challenges indoctrination'',
    ``Independent actions'': ``2 because the post emphasizes students making their own choice about whether or not to recite the pledge'',
    ``Stimulation'': ``X because the post does not concern itself with novelty, excitement, or variety in experience'',
    ``Personal security'': ``X because the post doesn't touch on an individual's need for personal safety or stability'',
    ``Societal security'': ``X because there is no focus on maintaining order or stability in society as a whole'',
    ``Tradition'': ``0 because the post challenges rather than supports traditional rituals like the pledge'',
    ``Lawfulness'': ``0 because the post promotes questioning authority and rules, rather than upholding them'',
    ``Respect'': ``0 because the post challenges the idea that children should show automatic respect for national symbols or authority'',
    ``Humility'': ``X because there is no emphasis on modesty or self-effacement; the tone is assertive and critical'',
    ``Responsibility'': ``X because the post does not promote fulfilling duties or obligations; it focuses on challenging them'',
    ``Caring'': ``1 because there is a weak presence of caring, shown through concern for the autonomy and mental freedom of children'',
    ``Equality'': ``2 because the post strongly promotes equality by highlighting systemic injustice and advocating for all individuals, including marginalized groups, to be represented fairly'', 
    ``Connection to nature'': ``X because the post makes no mention of the natural world or environmental concerns'',
    ``Tolerance'': ``1 because there is a weak presence of tolerance, seen in the critique of religious imposition and the desire for inclusion of different beliefs'' 
    }
}
    
\end{lstlisting}

\vspace{1.5em}

\verb|{example2}|

\begin{lstlisting}[  basicstyle=\ttfamily\scriptsize,
  breaklines=true,
  breakindent=0pt,
  breakautoindent=false,
  xleftmargin=0pt,
  numbers=none,
  columns=fullflexible] 
 { ``Rating'': {
     ``Reputation'': ``X because the post does not focus on social image or public approval'',
    ``Power'': ``X because the post does not promote dominance or control but rather transparency'',
    ``Wealth'': ``X because no reference to money or material status is made'',
    ``Achievement'': ``X because no emphasis on personal ambition or success'',
    ``Pleasure'': ``0 because the post does not reference fun or enjoyment'',
    ``Independent thoughts'': ``X because the post is not focused on wanting individuals to think for themselves more or be more creative'',
    ``Independent actions'': ``X because the post does not focus on wanting others to act independently or making their own decisions'',
    ``Stimulation'': ``X because no focus on novelty or excitement is present'',
    ``Personal security'': ``2 because strongly present due to the emphasis on protecting citizens and officers through transparent policing'',
    ``Societal security'': ``2 because strongly present because it highlights increased public safety and institutional stability through accountability'',
    ``Tradition'': ``X because the post does not focus on preserving traditional practices'',
    ``Lawfulness'': ``1 because somewhat present by endorsing lawful behavior and use of technology to uphold justice'',
    ``Respect'': ``X because no emphasis on respect toward authority or individuals'',
    ``Humility'': ``X because the post does not focus on modesty or self-effacement'',
    ``Responsibility'': ``2 because strongly present in encouraging accountability and duty from police officers'',
    ``Caring'': ``1 because weak presence as it shows concern for fairness and justice for all parties involved'',
    ``Equality'': ``2 because strongly present due to advocating for fair treatment of citizens and officers alike'',
    ``Connection to nature'': ``X because no mention or focus on nature or environment'',
    ``Tolerance'': ``X because does not focus on acceptance of different beliefs or lifestyles'' } }
    
\end{lstlisting}

    \vspace{1.5em}
    IMPORTANT FORMATTING NOTES: ...

\end{tcolorbox}

\end{document}